\documentclass[aps,amssymb,prb,showpacs,twocolumn]{revtex4}

\usepackage{epsf}
\usepackage{amsmath}
\usepackage{graphicx}

\begin{document}

\title{Decay in an asymmetric SQUID}

\author{J. P. Palomares-B\'{a}ez$^{1}$ and B. Ivlev$^{2}$} 

\affiliation
{$^{1}$Instituto Potosino de Investigaci\'{o}n Cientifica y Tecnol\'{o}gica, San Luis Potos\'{\i}, San Luis Potos\'{\i} 78231, Mexico\\
$^{2}$Instituto de F\'{\i}sica, Universidad Aut\'onoma de San Luis Potos\'{\i}\\
San Luis Potos\'{\i}, San Luis Potos\'{\i} 78000, Mexico}


\begin{abstract}
Quantum tunneling in an asymmetric (with strongly different capacitances) SQUID is studied. Since capacitances play a role of masses one phase, related to a large mass, becomes "heavy" and remains always a constant in a tunneling process. Tunneling in an asymmetric SQUID becomes one-dimensional with a condition of optimization of tunneling probability with respect to a value of
the "heavy" phase. An unusual temperature dependence of the tunneling probability is obtained. It has a finite slope at zero temperature and a transition between thermally assisted tunneling 
and pure activation can be not smooth depending on current through a SQUID.

\end{abstract} \vskip 1.0cm


\maketitle

\section{INTRODUCTION}
\label{intr}
A decay of a zero voltage state via quantum tunneling of a phase across a potential barrier in Josephson junctions is possible.\cite{LEGGETT1,OVC1,CLARKE1,CLARKE2} Tunneling in a single Josephson junction is similar to a conventional one-dimensional quantum mechanical process. In this case the tunneling mechanism is described by theory of Wentzel, Kramers, and Brillouin (WKB). \cite{LANDAU} Tunneling occurs from a classically allowed region which is a conventional potential well where energy levels are quantized. \cite{CLARKE3,CLARKE4,UST1,UST2} Quantum coherence between potential wells was demonstrated.\cite{LEGGETT2,NAK,TOLP,MOO}

Besides single Josephson junctions, superconducting quantum interference devices (SQUID) are also a matter of active investigation. \cite{CHEN,OVC2,CRIST,WANG,BAL,CAST,MITR,BLAT,IVLEV} A SQUID consists of two junctions and, therefore, represents a two-dimensional system where macroscopic quantum tunneling is also possible.

Tunneling in multi-dimensional systems was described in literature. \cite{COLEMAN1,COLEMAN2,SCHMID1,SCHMID2,MELN} According to those theories, there is a certain underbarrier path where a wave function is localized and it decays along the path. However, a tunneling scenario in two (many) dimensions can substantially differ from that main-path mechanism. For example, in a SQUID a main path can split by two ones \cite{OVC2} and even by an infinite number of equivalent paths which interfer providing multi-path tunneling. \cite{IVLEV} See  also \cite{IVLEV1,IVLEV2,DYK}.

Capacitances of two junctions in a SQUID play a role of masses. When one capacitance is large the masses become very different and one phase becomes "heavy". This provides an additional interest for study of an asymmetric SQUID. A behavior of an asymmetric SQUID after tunneling was investigated.\cite{BLAT} In this paper we also focus on strongly asymmetric SQUIDs, namely, 
on tunneling process in them.

During tunneling process a motion along the "heavy" coordinate is weakly generated and the process of barrier crossing becomes almost one-dimensional when the "heavy" phase is fixed. This fixed value should be determined from a condition of maximum of a tunneling probability.

That program has been performed in the paper. There are two unusual features of results.

First, the tunneling probability, as a function of temperature, has a finite slope at low temperature. This contrasts to a temperature dependence for a one-dimensional barrier where that slope is zero.

Second, a transition at a finite temperature between thermally assisted tunneling and pure activation changes its character when current approaches the critical value. At those currents temperature dependence of tunneling probability exhibits a finite jump of slopes at the transition temperature. When current is not too close to the critical value the transition is smooth as for a one dimensional barrier.

In Sec.~\ref{gen} we apply to a SQUID a semiclassical formalism of Hamilton-Jacobi. In Sec.~\ref{probab} the method of classical trajectories in imaginary time is used which accounts for an
optimization of tunneling probability with respect to a value of the "heavy" phase. In Sec.~\ref{disc} it is argued that experimental observations of the proposed phenomena in SQUID is real.
\section{FORMULATION OF THE PROBLEM}
\label{formulation}
We consider a dc SQUID, consisting of two Josephson junctions with phases $\varphi_{1}$ and $\varphi_{2}$, with no dissipation when the two junctions are inductively coupled. Critical currents of the junctions are equal but capacitances, $C_{1}$ and $C_{2}$, are different so that
\begin{equation}
\label{0}
M=\frac{C_{2}}{C_{1}}\gg 1.
\end{equation}
A classical behavior of phases corresponds to conservation of the total energy
\begin{eqnarray}
\label{1}
\nonumber
&&E_{0}=\frac{E_{J}}{2\omega^{2}}\left[\left(\frac{\partial\varphi_{1}}{\partial t}\right)^{2}+
M\left(\frac{\partial\varphi_{2}}{\partial t}\right)^{2}\right]+E_{J}\bigg[-\cos\varphi_{1}\\
&&-\cos\varphi_{2}-j(\varphi_{1}+\varphi_{2})+\frac{1}{2\beta}(\varphi_{1}-\varphi_{2})^{2}\bigg],
\end{eqnarray}
where the dimensionless current $j=I/2I_{c}$, the Josephson energy $E_{J}=\hbar I_{c}/2e$, the plasma frequency $\omega=\sqrt{2eI_{c}/\hbar C_{1}}$, and the coupling parameter $\beta=2\pi LI_{c}/\Phi_{0}$ are introduced. Here $I_{c}$ and $L$ are critical current and inductance of each individual junction. The magnetic flux quantum is $\Phi_{0}=\pi\hbar c/e$. As follows from Eqs.~(\ref{0}) and (\ref{1}), $\varphi_{2}$ is a "heavy" phase.
\begin{figure}
\includegraphics[width=6.5cm]{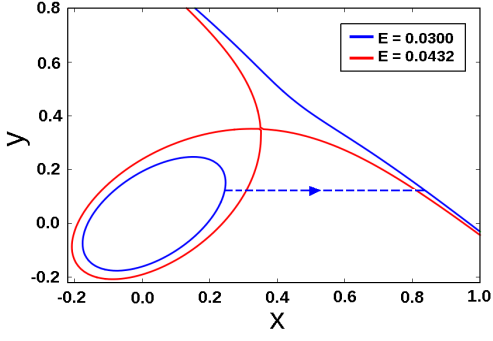}
\caption{\label{fig1}(Color online) Curves of a constant energy $V(x,y)=E$ at $\alpha=0.9$. Tunneling occurs along the dashed line, $y=y_{0}$, where an underbarrier wave function is localized.}
\end{figure}

Below we consider a large $\beta$ and the total current $I$ close to its critical value, $(1-j)\ll 1$. New variables are introduced by the relations
\begin{eqnarray}
\label{3}
&&\varphi_{1}=\frac{\pi}{2}+(3x-1)\sqrt{2(1-j)}+\frac{3x}{\beta}\\
\nonumber
&&\varphi_{2}=\frac{\pi}{2}+(3y-1)\sqrt{2(1-j)}+\frac{3y}{\beta}.
\end{eqnarray}
Below time is measured in the unit of
\begin{equation}
\label{3a}
t_{0}=\frac{\sqrt{2\beta}}{\omega}\sqrt{\frac{\alpha}{1+\alpha}},
\end{equation}
where the coupling parameter is
\begin{equation}
\label{4}
\alpha=\frac{1}{\beta\sqrt{2(1-j)}}.
\end{equation}
The energy (\ref{1}) takes the form
\begin{equation}
\label{5}
E_{0}=\frac{\hbar B}{t_{0}}\left[\frac{1}{2}\left(\frac{\partial x}{\partial t}\right)^{2}+
\frac{M}{2}\left(\frac{\partial y}{\partial t}\right)^{2}+V(x,y)\right],
\end{equation}
where
\begin{equation}
\label{6}
B=\frac{B_{0}}{\sqrt{2}}\left(\frac{1+\alpha}{\alpha}\right)^{5/2},\hspace{0.5cm}B_{0}=\frac{9E_{J}}{\hbar\omega\beta^{5/2}}.
\end{equation}

The potential energy is
\begin{equation}
\label{7}
V(x,y)=V_{0}(x)+V_{0}(y)-\frac{2\alpha xy}{1+\alpha},
\end{equation}
where $V_{0}(x)=x^{2}-x^{3}$. $B$ in Eq.~(\ref{5}) is called semiclassical parameter. When $B$ is large the phase dynamics is mainly classical. Below we consider that case, $1\ll B$.
\begin{figure}
\includegraphics[width=5.5cm]{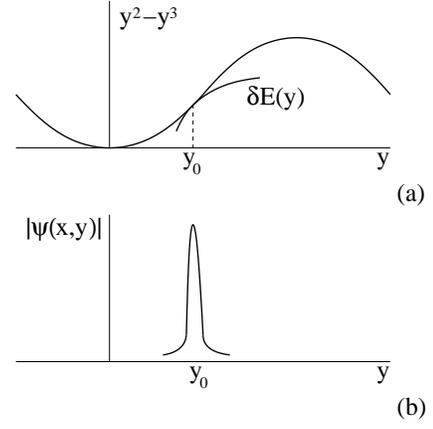}
\caption{\label{fig2} (a) Form of $\delta E(y)$. (b) Corresponding density distribution is Gaussian. It is plotted for some $x$ under the barrier.}
\end{figure}
\section{DESCRIPTION OF TUNNELING}
\label{gen}
A classical dynamics of phases in a SQUID relates to Eqs.~(\ref{5}) and (\ref{7}). The effective particle moves in the classically allowed region, in a vicinity of the point $x=y=0$, which is restricted by the potential barrier. As known, the particle can tunnel through the barrier resulting in experimentally observable phase jumps. Character of tunneling
depends on coupling strength $\alpha$ between the two junctions. At $\alpha=0.90$ the curves of equal potential, $V(x,y)=E$, are shown in Fig.~\ref{fig1}. An effective particle tunnels from
one classically allowed region (the potential well) to another (the outer region).

To quantitatively study the problem of two-dimensional tunneling one should solve the Schr\"{o}dinger equation with the exact potential (\ref{7}). Since the potential barrier is almost classical one can apply a semiclassical method when a wave function has the form
\begin{equation}
\label{8}
\psi\sim\exp(iB\sigma),
\end{equation}
where the classical action is $\hbar B\sigma$. $\sigma$ satisfies the equation of Hamilton-Jacobi \cite{LANDAU}
\begin{equation}
\label{9}
\frac{1}{2}\left(\frac{\partial\sigma}{\partial x}\right)^{2}+\frac{1}{2M}\left(\frac{\partial\sigma}{\partial y}\right)^{2}+V(x,y)=E.
\end{equation}
We define the energy $E$ by the relation
\begin{equation}
\label{9a}
E_{0}=\frac{\hbar B}{t_{0}}.
\end{equation}
At a large $M$ a solution of Eq.~(\ref{9}) can be written in the form $\sigma=\sigma_{0}+\sigma_{1}$ where $\sigma_{1}$ is small and $\sigma_{0}$ is given by
\begin{eqnarray}
\nonumber
\frac{\sigma_{0}(x,y)}{\sqrt{2}}=i\int^{x}dx_{1}\sqrt{x^{2}_{1}-x^{3}_{1}-\frac{2\alpha x_{1}y}{1+\alpha}-E+\delta E(x_{1},y)}\\
\label{9b}
+\sqrt{M}\int^{y}dy_{1}\sqrt{\delta E(x,y_{1})-y^{2}_{1}+y^{3}_{1}},~~~~~~~~~~~~~~
\end{eqnarray}
where $\delta E(x,y)$ is some function to be specified. It is easy to conclude that the correction $\sigma _{1}$ is small (proportional to $1/\sqrt{M}$) when the derivative $\partial\delta E/\partial x$ is small (proportional to $1/M$). So we consider below $\delta E(y)$.
\begin{figure}
\includegraphics[width=5.5cm]{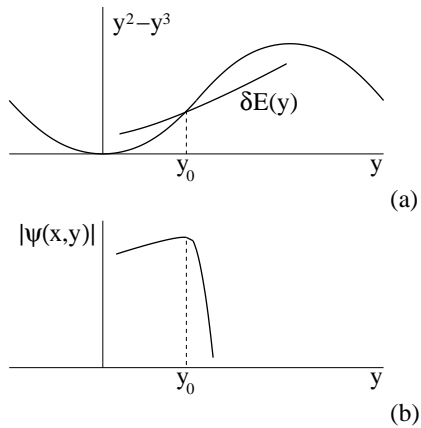}
\caption{\label{fig3} (a) Form of $\delta E(y)$. (b) Corresponding density distribution, plotted for some $x$ under the barrier, is not Gaussian. See the text.}
\end{figure}

The function $\delta E(y)$ is determined by a state in the well from which tunneling occurs.

When the case of Fig.~\ref{fig2}(a) is realized the last term in Eq.~(\ref{9b}) provides a Gaussian distribution of density around the line $y=y_{0}$ shown in Fig.~\ref{fig2}(b) for some $x$ under the barrier. This is analogous to a conventional scenario of tunneling in two dimensions. \cite{SCHMID1,SCHMID2}

In the case of $\delta E(y)$ of Fig.~\ref{fig3}(a) at $y_{0}<y$ the density drops down  under the barrier due to the second term in Eq.~(\ref{9b}) and at $y<y_{0}$ due to the first one. Therefore the underbarrier density in that case is also localized in a vicinity of the line $y=y_{0}$. It is shown in Fig.~\ref{fig3}(b) for some $x$ under the barrier. This is not a Gaussian distribution but one of the type $\exp[-c(y-y_{0})^{3/2}]$. At $y_{0}<y$ the parameter $c\sim\sqrt{M}$ is large.

The situations in Figs.~\ref{fig2} and \ref{fig3} relate to different types of states in the potential well from which tunneling occurs. In the case of Fig.~\ref{fig2} a Gaussian distribution of density holds also in the well because the last term in  Eq.~(\ref{9b}) dominates. In the case of Fig.~\ref{fig3} the distribution in the well is analogous to Fig.~\ref{fig3}(b) when the part at $y<y_{0}$ is horizontal and, therefore, the state is distributed over a finite distance $y$ in the well.

In the both cases, Figs.~\ref{fig2} and \ref{fig3}, tunneling occurs along the certain line $y=y_{0}$ and $y_{0}$ should be determined from the condition of maximum of a tunneling probability. This corresponds to classical mechanics when a particle does not move along a "heavy" direction. With an exponential accuracy a tunneling probability is the same for the both types of states in the well.
\section{TUNNELING PROBABILITY}
\label{probab}
Since tunneling occurs along the line $y=y_{0}$ one can use a WKB approach as in a one-dimensional case. The probability of tunneling with a fixed energy $E$ is
\begin{equation}
\label{10}
\Gamma(E)\sim\exp[-2BA(E)],
\end{equation}
where
\begin{equation}
\label{11}
A(E)=\sqrt{2}\int dx\sqrt{v(x)-E}.
\end{equation}
The one-dimensional potential $v(x)$ is given by
\begin{equation}
\label{11a}
v(x)=x^2-x^3-\frac{2\alpha y_{0}}{1+\alpha}x+y^{2}_{0}-y^{3}_{0}.
\end{equation}
\begin{figure}
\includegraphics[width=7cm]{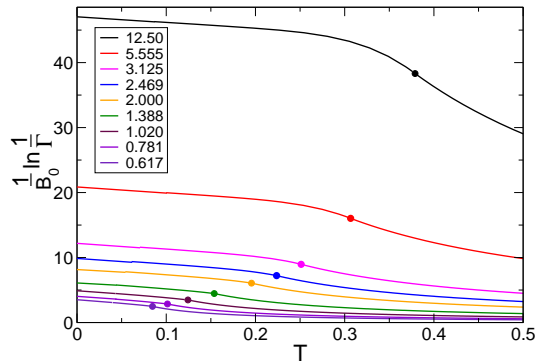}
\caption{\label{fig4}(Color online) $\Gamma$ is tunneling probability and temperature $T$ is measured in the units of $\hbar\omega/\sqrt{\beta}$. The numbers mark values of the parameter $(1-j)\beta^{2}$.}
\end{figure}
The integration in Eq.~(\ref{11}) is restricted by the classically forbidden region where $E<v(x)$.

Tunneling probability at a fixed temperature accounts for the Gibbs factor and is determined by
\begin{equation}
\label{11b}
\Gamma\sim\exp\left[-2BA(E)-\frac{E_{0}}{T}\right],
\end{equation}
with a subsequent optimization with respect to $E$. Taking Eq.~(\ref{9a}), one can write Eq.~(\ref{11b}) in the form
\begin{equation}
\label{11c}
\Gamma\sim\exp\left(-2BA_{T}\right),
\end{equation}
where 
\begin{equation}
\label{11d}
A_{T}=A(E)+\frac{E}{\theta}.
\end{equation}
The parameter $\theta$ is connected with temperature
\begin{equation}
\label{12}
\theta=\frac{2T\sqrt{\beta}}{\hbar\omega}\sqrt{\frac{2\alpha}{1+\alpha}}.
\end{equation}

Minimization of $A_{T}$ with respect to energy defines $E$ by the equation
\begin{equation}
\label{13}
\frac{1}{\theta}=\frac{1}{\sqrt{2}}\int\frac{dx}{\sqrt{v(x)-E}}.
\end{equation}
The parameter $y_{0}$ should be chosen to minimize $A_{T}$.
\begin{figure}
\includegraphics[width=7cm]{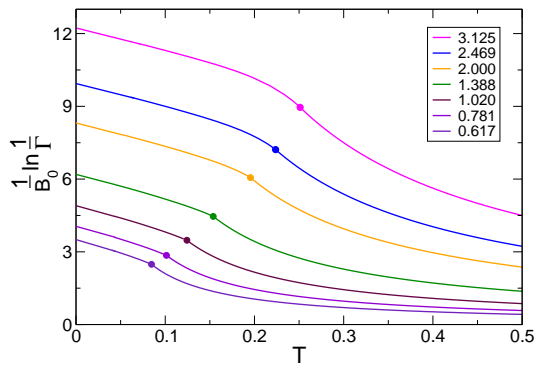}
\caption{\label{fig5}(Color online) Amplification of the lower set of curves in Fig.~\ref{fig4}. Left parts of the curves relate to a thermally assisted tunneling and right parts pertain to
pure activation.}
\end{figure}
By introducing imaginary time $t=i\tau$ the action $A_{T}$ can be written in the form
\begin{equation}
\label{18}
A_{T}=\int^{1/\theta}_{0}d\tau\left[\frac{1}{2}\left(\frac{\partial x}{\partial\tau}\right)^{2}+v(x)\right],
\end{equation}
where the classical trajectory under the barrier is determined by Newton's equation
\begin{equation}
\label{14}
\frac{\partial^{2} x}{\partial\tau^{2}}=2x-3x^{2}-\frac{2\alpha y_{0}}{1+\alpha}
\end{equation}
with zero velocities at the terminal points, $\tau=0$ and $\tau=1/\theta$. According to Eq.~(\ref{13}), $1/\theta$ is the underbarrier time of motion between two terminal points. In terms of trajectories, the condition of minimum $A_{T}$ with respect to $y_{0}$ takes the form
\begin{equation}
\label{17}
2y_{0}-3y^{2}_{0}=\frac{2\alpha\theta}{1+\alpha}\int^{1/\theta}_{0}xd\tau.
\end{equation}

For a strongly asymmetric SQUID, a large $M$, tunneling occurs along a straight line $y=y_{0}$ shown in Fig.~\ref{fig1}. The action (\ref{18}) depends on two parameters, $\alpha$ and $T\sqrt{\beta}/\hbar\omega$.

The tunneling probability satisfies the relation
\begin{equation}
\label{20}
\frac{1}{B_{0}}\ln\frac{1}{\Gamma}=\sqrt{2}\left(\frac{1+\alpha}{\alpha}\right)^{5/2}A_{T}\left(\alpha,T\sqrt{\beta}/\hbar\omega\right).
\end{equation}
A recipe of calculation of the action $A_{T}$ is the following. At fixed $\alpha$, $y_{0}$, and $\theta$ one should find a solution of Eq.~(\ref{14}) with zero velocities,
$\partial x/\partial\tau=0$, at $\tau=0,1/\theta$. That solution has to be inserted into the relation (\ref{17}) which defines $y_{0}$ at fixed $\theta$ and $\alpha$. The solution with the
defined $y_{0}$ should be substituted into Eq.~(\ref{18}) which produces $A_{T}\left(\alpha,T\sqrt{\beta}/\hbar\omega\right)$. We demonstrate in Sec.~\ref{low} how this scheme works in the case of low temperatures.
\section{TUNNELING AT LOW TEMPERATURES}
\label{low}
At low temperatures the energy $E$ should be close to the minimum of the potential $v(x)$ providing a long underbarrier time $1/\theta$. With the value of energy
\begin{equation}
\label{23}
E=\frac{1+2\alpha}{(1+\alpha)^{2}}y^{2}_{0}
\end{equation}
the action takes the form
\begin{equation}
\label{24}
A_{T}=\frac{4\sqrt{2}}{15}-\frac{2\sqrt{2}\alpha y_{0}}{1+\alpha}+\frac{E}{\theta}.
\end{equation}
A minimization with respect to $y_{0}$ of the action (\ref{24}), accounting for (\ref{23}), is equivalent to Eq.~(\ref{17}). The resulting action, at low dimensionless temperature $\theta$, is
\begin{equation}
\label{25}
A_{T}=\frac{4\sqrt{2}}{15}-\frac{2\alpha^{2}}{1+2\alpha}\hspace{0.5mm}\theta.
\end{equation}
\section{RESULTS}
\label{num}
We performed a numerical solution of Eq.~(\ref{14}). The results for the tunneling probability are presented in Figs.~\ref{fig4} and \ref{fig5} where temperature $T$ is
measured in the units of $\hbar\omega/\sqrt{\beta}$. Each curve in Figs.~\ref{fig4} and \ref{fig5} consists of two parts. To the left of a dot each curve relates to above trajectory
calculations corresponding to thermally assisted tunneling. To the right of a dot a curve is solely due to thermal activation
\begin{equation}
\label{26}
\frac{1}{B_{0}}\ln\frac{1}{\Gamma}=\frac{\hbar\omega}{T\sqrt{\beta}}\hspace{0.5mm}\frac{2}{27\alpha^{3}}
\begin{cases}
(1-\alpha)(1+2\alpha)^{2},&\alpha<1/2\\
2,&1/2<\alpha
\end{cases}
\end{equation}

The activation energy is given by the saddle point $V(x_{s},y_{s})$ which coinsides with the crossing point of two curves in Fig.~\ref{fig1}. The steepest descent in Fig.~\ref{fig1} goes along the direction $x=y$. The saddle point $\{x_{s},y_{s}\}$ is determined by the conditions $\partial V(x,y)/\partial x=\partial V(x,y)/\partial y=0$. At $1/2<\alpha$
\begin{equation}
\label{27}
x_{s}=y_{s}=\frac{2}{3(1+\alpha)}
\end{equation}
and at $\alpha<1/2$
\begin{equation}
\label{28}
x_{s},y_{s}=\frac{1+2\alpha\pm\sqrt{1-4\alpha^{2}}}{3(1+\alpha)}.
\end{equation}
\begin{figure}
\includegraphics[width=7cm]{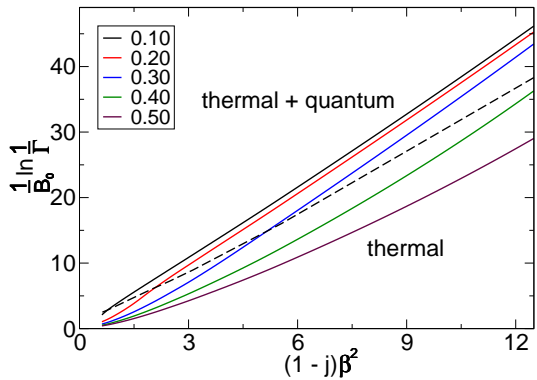}
\caption{\label{fig6}(Color online) Tunneling probability $\Gamma$ versus current for various values of the dimensionless temperature $T\sqrt{\beta}/\hbar\omega$. The dashed curve separates thermally assisted tunneling from a pure thermal activation.}
\end{figure}

At $\alpha<1/2$ the transition to a pure activation regime is smooth as in a one-dimensional case. It is analogous to type II phase transition. This corresponds to $2<(1-j)\beta^{2}$ in Figs.~\ref{fig4} and \ref{fig5}. At $1/2<\alpha$ the transition to the activation regime reminds type I phase transition. A derivative with respect to temperature jumps at those points which
can be observed in Fig.~\ref{fig5}.

Numerically calculated curves in Figs.~\ref{fig4} and \ref{fig5} match at low temperatures the analytical dependence followed from Eqs.~(\ref{20}) and (\ref{25}). At low temperatures
\begin{equation}
\label{29}
\frac{1}{B_{0}}\ln\frac{1}{\Gamma}=\frac{8}{15}\left(\frac{1+\alpha}{\alpha}\right)^{5/2}\left[1
-\frac{T\sqrt{\beta}}{\hbar\omega}\frac{15\alpha^{2}\sqrt{\alpha}}{(1+2\alpha)\sqrt{1+\alpha}}\right].
\end{equation}
We note that the slope in the temperature dependence (\ref{29}) is finite.

The tunneling probability $\Gamma$ as a function of the parameter $(1-j)\beta^{2}$ is plotted in Fig.~\ref{fig6} for different values of the dimensionless temperature $T\sqrt{\beta}/\hbar\omega$.
This plot shows how $\Gamma$ depends on current at a fixed temperature.
\section{DISCUSSIONS}
\label{disc}
Quantum tunneling across a one-dimensional static potential barrier is described by WKB theory. Accordingly, in two dimensions the main contribution to tuneling probability comes from an extreme path linking two classically allowed regions. The path is a classical trajectory with real coordinates which can be parametrized by imaginary time. The underbarrier trajectory is a solution of Newton's equation in imaginary time. Under the barrier the probability density reaches a maximum at each point of the trajectory along the orthogonal direction with respect to it. 
Along that direction the density has a Gaussian distribution. Therefore around the trajectory, which plays a role of a saddle point, quantum fluctuations are weak. The wave function, tracked along that trajectory under the barrier, exhibits an exponential decay analogous to WKB behavior. This constitutes a conventional scenario of tunneling in multi-dimensional case \cite{SCHMID1,SCHMID2} which can be called main-path tunneling.\cite{IVLEV}

In a symmetric (not very asymmetric) SQUID besides the conventional main-path tunneling \cite {OVC2} also multi-path tunneling is possible.\cite{IVLEV} In that case a density distribution
under a barrier is not as in Fig.~\ref{fig2}(b), but of the type as in Fig.~\ref{fig3}(b). For a very asymmetric SQUID, considered in this paper, the both mechanisms result in the same tunneling exponents since the underbarrier channel shrinks to the line $y=y_{0}$ due to the mass difference.

We used a semiclassical approximation when there are many levels in the well. This approach sometimes is not appropriate in one-dimensional Josephson junctions where a barrier is weakly
transparent but nevertheless there is only a few levels in the well (say, five).\cite{CLARKE1,CLARKE2,CLARKE3,CLARKE4,UST1} In a SQUID based on two such junctions the number of levels can
be roughly estimated as $5\times 5$. In our case of a strongly asymmetric SQUID that number should be multiplied by the large parameter $\sqrt{M}$. Therefore the approximation of a large
number of levels in the well is reasonable.

We propose two peculiarities of tunneling in an asymmetric SQUID which do not exist in a single junction and in a not very asymmetric SQUID.

One of them is temperature dependence of tunneling probability at low temperature. According to Eq.~(\ref{29}), the curves in Fig.~\ref{fig5} have a finite slope at low temperature. In one
dimension the slope is zero due to the exponent $\exp(-const/T)$ instead of $T$ in Eq.~(\ref{29}).

The second peculiarity is an unusual transition between thermally assisted tunneling and pure activation marked by dots in Fig.~\ref{fig5}. At $2<(1-j)\beta^{2}$ the transition is smooth but at
$(1-j)\beta^{2}<2$ there are jumps of slopes in Fig.~\ref{fig5}.

Low dissipation regime and parameters $M\simeq 35$ and $\beta\simeq 15$ correspond to reality in experiments with SQUIDs and fit the developed theory. It is more convenient in experiments to obtain a set of curves as in Fig.~\ref{fig6} since usually measurements are run at a fixed temperature. A dependence on temperature, as in Fig.~\ref{fig5}, also can be obtained. This would provide an experimental check of the predicted dependences on temperature and current 

\section{CONCLUSION}
Quantum tunneling in an asymmetric (with strongly different capacitances) SQUID is studied. Since capacitances play a role of masses one phase, related to a large mass, becomes "heavy" and remains always a constant in a tunneling process. Tunneling in an asymmetric SQUID becomes one-dimensional with a condition of optimization of tunneling probability with respect to a value of
the "heavy" phase. An unusual temperature dependence of the tunneling probability is obtained. It has a finite slope at zero temperature and a transition between thermally assisted tunneling 
and pure activation can be not smooth depending on current through a SQUID.

\acknowledgments
We thank G. Blatter, S. Butz, V. B. Geshkenbein, and A. V. Ustinov for valuable discussions.

\end{document}